# Choice of a Mentor: A Subjective Evaluation of Expectations, Experiences and Feedbacks

Kaibalyapati Mishra, Centre for Economic Studies & Policy, ISEC Bangalore – 560072

**Abstract:**

Recent trends in academics show an increase in enrollment levels in higher education. Predominantly in Doctoral programmes, where individual scholars, institutes and supervisors play the key roles. The human factor at receiving end of academic excellence is the scholar having a supervisor at the facilitating end. In this paper, I try to establish the role of different factors & availability of information about them in forming the basic choice set in a scholar's mind. After studying three different groups of individuals, who were subjected to substitutive choices, we found that scholars prefer an approachable, moderately intervening and frequently interacting professor as their guide.

**Introduction**:

Doctorate of Philosophy in any subject is widely believed to the highest degree of excellence in the same discipline. Among the crucial factors that affect and influences the scholar's performance, the role of a GOOD mentor (Supervisor / Guide) is vital. While the decision of whom to choose as the mentor is left to the scholar, several factors determine such a decision. In developing countries like India where the number of PhD admissions is increasing 10% per annum[1], while there prevails lack of capable human resource guidance in research, the choice of a mentor becomes a multivariate function of various individual, academic & professional attributes which are compositely subject to availability constraints of institutions. This project report analyses such distinctive factors that constitute the choice set of the scholar. This also tries to understand the nature of change in such behaviours over the length of time as the scholar moves from being an aspirant to an active chooser of mentor to a full-fledged scholar. This project report also elucidates the nature of various contradictory choices that individuals are subjected to due to the lack of availability of quality mentorship, individually and also concerning other factors of the choice set.

**Literature Review:**

The assumption of orthodox economics models, on the absolute rationality terms of individual sapiens are criticised as the foundation of behavioural economics. The ground stone of such an inglorious field has been kept on the fact that agents make their choices in a comprehensively inclusive context, which incorporates all the relevant details of the present situation, as well as expectations about all future opportunities and risks (Kanheman, 2003). Expectations come in different guises, with outcome expectations centred on prognostic beliefs about the consequences of engaging in treatment (Constantino, et.al. 2011). To predict choice behaviour, the standard practice of economists has been to infer decision processes from data on observed choices. When decision-makers act with partial information, economists typically assume that persons form probabilistic expectations for unknown quantities and maximize expected utility (Manski,2004). Such in this the decision-maker i.e. the aspiring scholar before the selection of her/his guide reserves some expectations in the form of attributes favourable to his/her doctoral objectives. Such expectations form a significant part of his/her influence that affects the choice of the mentor. This project report tries to incorporate the expectation of scholars from their potential mentor is being studied by analysing the behaviour of students who are on the verge of completing their Master's degree or are already enrolled in the PhD programme, but now undertaking coursework, after successful completion of which they will have to choose they mentors respectively.

Experiences are expectations' interaction with reality over the length of time and space. While sometimes they incentivize future outcomes by playing the role of directive role in decision making, they also form a critical role in framing future expectations as well (Mincer, 1974). The review of the effects of such nature experience on human cognitive function shows, how exposure to nature has been considered, and the role that individuals' preferences for nature may play in the impact of the environment on psychological functioning (Bratman, et.al. 2012). Similarly, contemporary views of individual behaviour in institutions stress that feedback is necessary for effective role performance, little attention is given to the psychological processes affected by it (IIgen et.al., 1979). The general human behaviour feedback formalism is developed (Evans,1985) in which the actual behavioural change in the choice system of a scholar is related to the intended or expected change of attributes over the distinctive time frames of previous to and after becoming a scholar.

**Conceptual Framework:**

The basic conceptual framework revolves around the possible characteristics of an ideal mentor. Following the general principles that govern individual intuition, it can be established that those characteristics can be classified into 5 major parts which are complementary to each other. Firstly, the expertise that an individual mentor has in the subject area of concern and the experience in terms of academic teaching and research matters the most. Secondly, the regularity of meeting and interaction and quality time disposal towards the growth of scholar by the mentor. Thirdly, feedback on constructing nature is also quite vital in guiding the investigating path pursued by the scholar. Advice and supports are fourthly prominent as these are the bare minimum expectations that the scholar usual has from his/her guide/mentor. Over and above the previously mentioned characteristics proper mediation and intervention along with the timely representation of the scholar also enhance the possibilities of exposer. These guiding characteristics are widely believed to be crucial determining factors individual expectation of the scholar as well.

The major concepts use in this report and their role & definitions are mentioned below.

| CONCEPT | ROLE / DEFINITION |
|---|---|
| EXPECTATION | The expectation here indicates the voluntary wishes of major characteristics of a mentor of a scholar. |
| EXPERIENCE | Following the same characteristics, the experience here denotes the real-time favourable and difficult situations faced while making the choice. |
| FEEDBACK | Feedback here is the impression of the choices, developed within the subject while occupying expectation, incurring experience and moving with them. |
| PEER INFLUENCE | It refers to the opinion given by the people (seniors other faculties inside and outside the institution) who are actively/inactively related or informed about the individual (mentor) or institution. |
| APPROACHABLENESS | This specific attribute refers to the fact the mentor maintains a positive motivation towards mentoring the scholar frequently. |

| RESOURCEFULNESS | It refers to the academic excellence of the mentor in terms of publications, interdisciplinary engagements, etc. |
|---|---|
| INTERACTION | This certainly indicates the number of times the mentor and student purposively interact to discuss matters of specific importance for the scholar's research. |
| INTERVENTION | It refers to the inclusion of the mentor's point of view in the scholar's research piece with or without his/her evident consent. |

**Objectives:** Following are the major objectives that this report tries to address

- To analyse factors influencing the scholar's choice behaviour.
- To understand changes in perception about mentorship in scholars.
- To understand the dynamics of influence non-human factors.

**Research Questions:** The systematic questions that the report wishes to check are the following:

- Are scholars inclined to adjust their research interest for the convenience of their mentors?
- If given a dichotomous choice of attributes like, approachableness and resourcefulness, which one will be more preferred?
- Whether frequency or duration of interaction with the guide matters more in a scholar's opinion?
- Does the peer influence as a factor contributing to the decision of scholar changes over the period?

**Data & Methodology:** The data here was collected through the questionnaire sent through a google form, to avoid the problem of reaching the subjects during this pandemic. The project followed a snowball sampling method to collect the data relevant to the study. The total sample was divided into three strata for which three different questionnaires were prepared. The first strata composed of students who have completed their Masters' degree or enrolled in a PhD programme, but are yet to choose their guides. The second strata composed of students, who have already enrolled and have just chosen their respective mentors/guides. The third and the last strata composed of scholars who are in the advanced stage of their research and about to submit their thesis. The institutional background was not a criterion for the selection of the samples into strata. The first strata were analysed to study the expectations, the second was studied to check for the experience and the last strata was used to examine the feedbacks of the respondents. Throughout the project simple algebraic and mathematical formulas of ratio, proportion and percentages are being used to analyse and compare the data sets.

A 31 number of observations could get collected for the first strata, while 22 and 13 observations were there in the second and third strata accordingly. An equal number of 33 respondents participated from both the sexes of male and female and the average age interval of respondents was between 25-35 years. Amongst the observation, 75% number of students either wished to study outside the state or were studying, while 87% of them were found to be interested in full-time research.

**Findings & Discussion:**

- **Findings from expectation analysis:** The study of expectation that comprised of individual students who are expecting to join a PhD programme, gave some interesting results. Over 97% of the observations here opined that the value of a good supervisor is more than that of the institution. The respondents however displayed their dissatisfaction related to the fact the information bouchers of institutes miss some very relevant information of the supervisors like Vacancy under them, Way of Guidance, Personal traits. However, several publications led to the attributes that dominated the good impression of a mentor. People believed that it is important to have a good thesis than to adjust the quality of the same. In this stage, most students (52%) believed that they can afford to adjust their research interest as per the convenience of their guide. Peer influence at the stage of formulating expectation mattered the most as 97% of the observations believed the same way. Most (58%) students believed that the approachableness of a mentor is more preferred over resourcefulness if the choice is exclusive. An important finding of the strata was that "Designation influences expectation" as students wanted Professor (58% of the observation) more than Associate Prof. (32%) than Asst. Prof. (19%). This goes against the prevailing narrative that assistant professors have more incentive to guide more students and the engagement provided by the mentors falls as we move up in the hierarchical structure. Based on interaction, in framing expectations, students believe that the frequency of interaction matters more than that of the duration.
- **Findings from experience analysis:** Here also a large (58%) number of respondents opined that the role of a mentor/guide is more than that of the supervisor. The respondents in this strata believed that availing a co-guide outside the institute faculty collegium will encourage better research results in case of interdisciplinary research. Here people claimed that they met their supervisors less than five times before actually making the choice. They believe that the choice after coursework leads to the better decision of the faculty. At this stage, 69% of the people believed that higher intervention is needed and in 46% of the cases peer influence mattered significantly.

- **Findings from the study of feedback:** Over 93% of respondents didn't wish to disclose their names here. While the role of supervisor was hailed over 70% of the observations, duration of quality interaction mattered in their opinion more than that of the frequency of the same. In all three stages number of publication has remained as the unique selling characteristics of a prospective mentor as compared to other ones. People here (56% of the obs.) believe that the resourcefulness of a mentor matters more than that of the approachableness of the guide. They also believed that the availability of more options leads to the better decision of supervisors.
- **A comparative analysis:** A relative analysis is being presented in the table (table 1.1) below. By studying through the rows we find that the feeling that supervisors play a vital role in that institute has decreased over the three stages, though it has dominated popular opinion. Similarly, approachable attitudes are believed to be more accurate in formulating expectations and experiences than that of the feedback case, which indicates that with time people weigh resourcefulness more. Intervention in the research work of a student is high in case of experience suggesting that maybe people at the initial stage of their research seek greater intervention which falls as they learn to do things by themselves in the later stages. Duration of quality interaction mattered in the later stages more as compared to frequency as suggested in the stage of formulation of expectation. However, peer influence has dominated in all three stages of study.

|  | **Expectation** | **Experience** | **Feedback** |
| --- | --- | --- | --- |
| Institute Vs Supervisor | 97% | 58% | 70% |
| Approachableness Vs Resourcefulness | Approachable | Approachable | Resourcefulness |
| Intervention | Moderate (58%) | High (69%) | Moderate (76%) |

| Interaction | Frequency (87%) | < 5times | Duration (47%) |
| --- | --- | --- | --- |
| Peer Influence | 98% | 46% | 77% |

Table. 1.1 – Expectation, experience and Feedback comparison table(* here % indicates the highest percentage of observation )

**Scope**: This idea is a pretty novel one. It can be useful to formulate policies for R & D in countries as a significant part of initial research is done by scholars only. In such cases, this can help us to better understand the scholar behaviour and such findings can be very helpful in making policies for higher education and research. It can be used to develop a Mentor Assessment Index that will assess and analyse the competence of a mentor in guiding students and the scores of such an index can be used for employment and promotion of faculties at different levels.

**Limitations & Conclusion:** The basic limitation of the study is that the objectives strongly demand a longitudinal study of selected and specific observations in all three stages of expectation formulation, experience and feedbacks. Another problem was that doing a subjective evaluation through google form is less appropriate as it is impossible to assess the behavioural reactions of them and here the study operates on an assumption that, respondents are responding honestly, however no such mechanism for its validation is available. To conclude, this can be asserted that this new area of study requires special focus and here exists enough importance for policies focused on higher education and research. The wide scope of the study can be extended further towards improving the research atmosphere at the individual level and at the institutional level too.